\documentclass[manuscript,screen]{acmart}

\usepackage{graphicx}
\usepackage{soul}
\usepackage{makecell}

\AtBeginDocument{%
  \providecommand\BibTeX{{%
    \normalfont B\kern-0.5em{\scshape i\kern-0.25em b}\kern-0.8em\TeX}}}

\setcopyright{acmcopyright}
\copyrightyear{2022}
\acmYear{2022}
\setcopyright{rightsretained}
\acmConference[CHI '22 Extended Abstracts]{CHI Conference on Human Factors in Computing Systems Extended Abstracts}{April 29-May 5, 2022}{New Orleans, LA, USA}
\acmBooktitle{CHI Conference on Human Factors in Computing Systems Extended Abstracts (CHI '22 Extended Abstracts), April 29-May 5, 2022, New Orleans, LA, USA}
\acmDOI{10.1145/3491101.3503733}
\acmISBN{978-1-4503-9156-6/22/04}

\begin{document}


\title[QTBIPOC PD]{QTBIPOC PD: Exploring the Intersections of Race, Gender, and Sexual Orientation in Participatory Design}

\author{Naba Rizvi}
\email{nrizvi@ucsd.edu}
\orcid{0000-0002-2777-700X}
\affiliation{%
  \institution{University of California, San Diego}
  \city{San Diego}
  \country{USA}
}
\author{Reggie Casanova-Perez, MS}
\email{reggiecp@uw.edu}
\orcid{0000-0002-8062-2947}
\affiliation{%
  \institution{University of Washington}
  \city{Seattle}
  \country{USA}
}

\author{Harshini Ramaswamy}
\orcid{0000-0002-5552-4579}
\affiliation{%
  \institution{University of California, San Diego}
  \city{San Diego}
  \country{USA}
}
\email{hramaswamy@ucsd.edu}

\author{Lisa G. Dirks, MS, MLIS}
\orcid{0000-0001-6768-6500}
\affiliation{%
  \institution{University of Washington}
  \city{Seattle}
  \country{USA}
}
\email{lgdirks@uw.edu}

\author{Emily Bascom}
\orcid{0000-0002-5140-6567}
\affiliation{%
  \institution{University of Washington}
  \city{Seattle}
  \country{USA}
}
\email{embascom@uw.edu}

\author{Nadir Weibel, PhD}
\orcid{0000-0002-3457-4227}
\affiliation{%
  \institution{University of California, San Diego}
  \city{San Diego}
  \country{USA}
}
\email{weibel@ucsd.edu}

\renewcommand{\shortauthors}{Rizvi and Casanova-Perez, et al.}

\begin{abstract}
  As Human-Computer Interaction (HCI) research aims to be inclusive and representative of many marginalized identities, there is still a lack of available literature and research on intersectional considerations of race, gender, and sexual orientation, especially when it comes to participatory design. We aim to create a space to generate community recommendations for effectively and appropriately engaging Queer, Transgender, Black, Indigenous, People of Color (QTBIPOC) populations in participatory design, and discuss methods of dissemination for recommendations. Workshop participants will engage with critical race theory, queer theory, and feminist theory to reflect on current exclusionary HCI and participatory design methods and practices.
\end{abstract}

\ccsdesc[500]{Human-centered computing~Participatory design}

\maketitle

\section{Background}

The power inequalities in HCI research involving marginalized communities have been investigated by researchers utilizing feminism~\cite{bardzell2011towards}, critical race theory~\cite{ogbonnaya2020critical}, decolonial theory~\cite{wong2020reflections, alvarado2021decolonial, cannanure2021decolonizing}, and queer theory~\cite{spiel2019queer}. While the intersectionality \cite{crenshaw1990mapping} of race, gender, sexual orientation, and gender identity has been explored in domains like law~\cite{crenshaw2015intersectionality}, healthcare~\cite{douglas2019black}, and education~\cite{schey2021race}, little work has been done on engaging with Queer, Transgender, Black, Indigenous, People of Color (QTBIPOC) populations in participatory design research. 

Participatory design is a well-known and often utilized method in HCI research~\cite{schuler1993participatory}, that typically aims to involve end-users and other stakeholders (family member, employees, customers, partners, etc.) in the design process. Harrington et al. describe participatory design as a powerful tool for involving marginalized communities in design~\cite{harrington2019deconstructing}. However, to employ inclusive research practices requires that researchers consider the social and cultural expectations associated with participatory design methods and how they will impact the communities involved in the design process, as well as the conventions for representing these participants when reporting on research, which remain unexplored for the QTBIPOC community.  

In this workshop we focus specifically on how to effectively incorporate intersectionality in participatory design. The central goal of intersectionality is to increase the understanding of how an individual’s social and political identities interact to influence their lived experiences~\cite{schlesinger2017intersectional}.  Incorporating intersectionality in participatory design research will help advance our understanding of a participant’s overlapping identities which inform their engagement with technology and their environment. Representing the intersectionality of participants' identities and experiences in research and design in participatory design is imperative to better serve marginalized communities, especially at the intersection of the Queer-Trans (QT), Black, Indigenous, and People of Color (BIPOC) communities. 

The main structure of our workshop is being developed and facilitated by a core group of organizers, supported by a steering committee that will help guide discussions on the intersectionality of race, gender, and sexual orientation in participatory design research. This committee includes researchers who have engaged in participatory design with QTBIPOC populations and community champions working at this intersection. 

Throughout the workshop we aim to create opportunities for guided discussions that will culminate in three distinct dissemination methods: (1) a website that extends the workshop into a living venue where researchers and community can consult results of the workshop and further engage in the future, (2) a public forum (e.g. a Medium blog) that summarizes our discussions on more inclusive categorizations of race, gender, and sexual orientation in research studies, and elicits further and broader community discussion beyond the scope of our workshop, and (3) a research paper outlining recommendations for engaging in participatory design with QTBIPOC populations that can help the HCI research community. Our workshop also provides researchers interested in engaging QTBIPOC people in HCI research an opportunity to network and form a community. 

\section{Organizers}

The organizers for this workshop have diverse backgrounds. Relevant to this workshop, we disclose that we have queer and international organizers from various backgrounds who are from countries in Asia, North America, South America, and Europe. This group of organizers includes individuals who have experience engaging in participatory design research with QTBIPOC populations.

\textbf{Naba Rizvi} (she/her), is a queer international student from Pakistan and Saudi Arabia, a 2nd year Ph.D student at the University of California, San Diego, and a 2020 NCWIT Collegiate Award recipient. Her research interests include exploring the intersections of race, gender, disability, and sexual orientation to guide inclusive participatory design research.

\textbf{Reggie Casanova-Perez, MS} (he/him), is a migrant trans non-binary 4th year Ph.D student in Biomedical and Health Informatics at University of Washington. His research interests focuses on making healthcare information systems more inclusive to gender-diverse patients using feminism and queer theory approaches.

\textbf{Harshini Ramaswamy} (she/her), is a 4th year South Asian American undergraduate student studying Cognitive Science with a specialization in Design and Interaction at the University of California, San Diego. Her research interests include human-centered design and research in technology and healthcare settings.

\textbf{Lisa G. Dirks, MS, MLIS}  (she/her), is a Ph.D. candidate at the University of Washington Information School. She is Alaska Native and White and grew up in a rural Unangan (Aleut) community in the Western Aleutian Islands. A social and information scientist, her research interests include health informatics, human-information interaction, user experience, Indigenous health, research dissemination, and community-engaged and participatory research.

\textbf{Emily Bascom} (she/her), is a 4th year undergraduate student studying Informatics with a specialization in Human-Computer Interaction at the University of Washington. Her research interests surround how information is created, disseminated, and uptaken by the public and how to holistically represent users in HCI research and informatics systems.

\textbf{Nadir Weibel, PhD} (he/him), is an Associate Professor in the Department of Computer Science and Engineering at UC San Diego, and a Research Health Science Specialist at the VA San Diego Health System. He is one of the faculty of the UCSD Design Lab, Contextual Robotics Institute, Center for Population and Health Systems, and the Research Center on Optimal Digital Ethics. His work on Human-centered eXtended Intelligence is situated at the intersection of computer science, design, and the health sciences, and strive to support a diverse end-user population through user-centered and participatory design.

\section{Website}
The workshop's website (\url{http://qtbipoc-hci.org}) will contain all information about our workshop, including how we solicit submissions. We include a Call for Participation, information about the organizers, the workshop's schedule and outline, the list of accepted submissions and post-workshop documents, and contact information. The website also serves as a repository for materials for our reading group and post-workshop plans, including public outreach updates. Accepted submissions regarding researchers’ experience engaging with the QTBIPOC community will be available through our website with their authors' consent. 

\section{Pre-Workshop Plans}
We will promote our workshop through our website and social media channels (Facebook groups, Twitter, institutional Slack channels, listservs, etc.). To broaden our pool of participants, we will also advertise the workshop with relevant organizations that have contact with the QTBIPOC community. Accepted submissions will be notified by March 20, 2022. Our submission form will include a section for participants to state their accessibility needs so we can provide accommodations as needed.

We will have two reading groups between the time when participants are notified that they have been accepted and the workshop date. These reading groups will require attendance at a one-hour remote session dedicated to discussing readings sent to the participants beforehand to prepare participants for the types of conversations that will be facilitated during the workshop. Our steering committee will help with selecting appropriate reading and resources. The goal of these conversations is to create recommendations for considering the intersectionality of race, gender, and sexual orientation in participatory design research. Suggested readings and resources will include peer-reviewed literature and works published through other mediums such as personal documents, Medium blogs, Ted Talks, and Zines. The spectrum of source types will provide a deeper, holistic understanding of both research and personal experiences related to this important subject matter. Transcripts for the meetings will be provided to all participants for accessibility and to review following the meeting.

\section{In-Person, Hybrid, or Virtual-Only}
Our workshop participants will attend two video conferencing calls before the workshop; each will be a 1-hour pre-workshop reading group discussion. During our workshop at CHI 2022, participants will have the option to either attend in-person or conference call in to the meeting. Additionally, attendees will be able to access all written materials used for the workshops, including recordings and transcripts from our reading groups, on our website following each meeting. We will provide accommodations as specified and requested by participants in the submissions form.

Since our goal is to engage a new community interested in HCI research with QTBIPOC people, we will open up a hybrid (in-person or remote) workshop participation to additional participants attending the CHI 2022 conference. We hope that additional participants will be able to learn and contribute to the workshop topic, and join the burgeoning community.

\section{Asynchronous Engagement}
To facilitate asynchronous discussions before the workshop, we will set up communication channels that are easy to join for our community participants (e.g. through Slack, Discord, or Discourse). Additionally, all reading materials, summaries, and notes from our discussions will be available to all participants through our website. During online engagements, we will offer live captioning and ensure our slideshow and other materials in the workshop follow the W3C guidelines for accessibility.

\section{Workshop Structure}

Table~\ref{tab:pre-workshop} shows our pre-workshop activities, while Table~\ref{tab:workshop} displays the proposed schedule for our workshop. Time will be allocated at the beginning of the workshop to make brief introductions, elaborate on community guidelines and expectations for participation in the workshop, and develop rapport with participants to facilitate open discussion surrounding the intersectionality of race, gender, and sexual orientation in HCI and design research. 

We will provide a summary of the motivations, goals and desired outcomes of the workshop. We will invite community champions and advocates working on intersectionality, race, gender, and sexual orientation to share their thoughts in short opening keynotes. We will then share the outcomes of our reading groups, and invite participants to share their perspectives on current or emerging challenges to be addressed in participatory design research in relation to intersectionality. We will also include an overview of anti-racist, queer, and feminist theories. We will then encourage discussion with participants on whether current standards for inclusive HCI methodologies and practices incorporate anti-racism, anti-queerphobia, and feminist theory. 


\begin{table}[h!]
\centering
\begin{tabular}{|l|l|}
\hline
\textbf{Period} & \textbf{Activity }         
\\\hline
January-February 2022    & Steering Committee and Organizers select readings and materials
\\\hline
March 20th & Reading Assignments are sent to Participants
\\\hline
4 weeks prior to CHI 2022 Workshop & 1st online reading group meet-up: Race in HCI
\\\hline
2 weeks prior to CHI 2022 Workshop & 2nd online reading group meet-up: Gender and Sexual Orientation in HCI
\\\hline
\end{tabular}%
\caption{Pre-Workshop Activities}
\label{tab:pre-workshop}

\bigskip

\begin{tabular}{|l|l|}
\hline
\textbf{Time (PST)}        & \textbf{Activity }
\\ \hline
9 am - 9.30 am    & Introductions \& Review of Workshop Goals          \\ \hline
9.30 am - 10.15 am    & Invited Community Keynotes                   
\\ \hline
10.15 am - 11.00 am   & \textbf{Summary of Reading Group Discussions:}\\ & -Race in HCI \\ & -Gender and sexual orientation in HCI
\\\hline
11.00 am - 11.30 am  & Break
\\\hline
11.30 am - 12:15 pm   & \textbf{Problems and Opportunities Discussion:} \\ & -Problems with design thinking methodology for the QTBIPOC community \\ & -How would HCI would benefit from using an intersectional approach 
\\\hline
12:15 pm - 12.45 pm    & Network and Discover  
\\\hline
12:45 pm - 1:45 pm    & Lunch Break                   \\\hline
1.45 pm - 3:30 PM & \textbf{Recommendations and Considerations Discussion:}\\ & \makecell[l]{-Recommendation generation for community-based participatory design with the\\ QTBIPOC community.} \\ &  \makecell[l]{-Brainstorming for instrument development that is respectful of people’s race, gender and\\ sexual orientation, and their intersections.}\\\hline
3.30 pm - 4 pm & Closing\\ 
\hline
\end{tabular}
\caption{Our proposed workshop schedule. Updated versions will be available on our website.}\footnote{https://www.qtbipoc-pd.org/schedule}
\label{tab:workshop}
\end{table}

\section{Post-Workshop Plans}

After the workshop, we will compile all material from the reading groups and the workshop discussions to produce: (1) a website making workshop resources available to the QTBIPOC community at large, (2) a Medium blog article for the general public, and (3) a research paper for the academic community.

We expect to release recommendations on how to consider intersectionality in participatory design research in the planned journal paper, and we will transform our workshop website into a living community page where we will disseminate our recommendations to interested community and research partners.

In addition to our dissemination plans, we expect that the workshop will attract QTBIPOC researchers and offer networking opportunities within this community. We will nurture these networking opportunities and facilitate them through the participants' chosen communication channel such as Slack or Discord.

\section{Call for Participation}
We invite participants interested in engaging with the intersection of race, gender, and sexual orientation in participatory design. Our workshop seeks to provide a safe space for scholars to share their experiences with participatory design methodology that excludes the experiences of marginalized participants. We have two goals: (1) to create recommendations for engaging in participatory design research with the QTBIPOC community, and (2) determine methods to disseminate the recommendations and reflections from the workshop.

Participants can join our workshop in two different ways:
\begin{itemize}
    \item \textbf{Positionality Submissions} - Emerging QTBIPOC researchers who want to contribute to the discussion can submit a positionality statement explaining what being QTBIPOC means to them, challenges they may have encountered as a QTBIPOC researcher, and how that has influenced their research interests.
    \item \textbf{Experience Submissions} - For researchers that have experience working with QTBIPOC communities who may or may not be QTBIPOC themselves. This submission will include a detailed explanation on their experience working with LGBTQ+ and/or BIPOC communities. The submissions do not need to focus on the intersection of the two identities.
\end{itemize}

Submission deadline is February 24th 2022 (23:59 PT). Positionality Submissions will not need to follow a specific format, whereas the Experience Submissions must follow the standard CHI format. All submissions must be sent as a PDF file using the Google Form available on the website.

Participants will be selected on the basis of the quality of their submissions. At least one author of each accepted submission must attend the workshop. The organizing committee can be contacted via email to answer relevant questions.


\bibliographystyle{ACM-Reference-Format}
\bibliography{QTBIPOC-PD_CHI2022}

\end{document}